\documentclass[%
reprint,
superscriptaddress,
amsmath,
amssymb,
aps,
floatfix,
]{revtex4-2}

\usepackage{graphicx}
\usepackage{units}
\usepackage{braket}
\usepackage{float}
\usepackage{csquotes}
\usepackage{array}

\usepackage[colorlinks, linkcolor=blue, citecolor=blue, urlcolor=blue, breaklinks=true]{hyperref}

\usepackage[dvipsnames]{xcolor}
\usepackage[normalem]{ulem}

\begin{document}

\title{Distribution of genuine high-dimensional entanglement over 10.2~km of noisy metropolitan atmosphere}

\author{Lukas Bulla}
\email{lukas.bulla@oeaw.ac.at}
\affiliation{Institute for Quantum Optics and Quantum Information (IQOQI), Austrian Academy of Sciences, Boltzmanngasse 3, 1090 Vienna, Austria}
\affiliation{Vienna Center for Quantum Science and Technology (VCQ), Faculty of Physics, University of Vienna, Boltzmanngasse 5, 1090 Vienna, Austria}

\author{Kristian Hjorth}
\affiliation{Institute for Quantum Optics and Quantum Information (IQOQI), Austrian Academy of Sciences, Boltzmanngasse 3, 1090 Vienna, Austria}
\affiliation{ 
Department of Physics, Faculty of Natural Sciences, Norwegian University of Science and Technology (NTNU), NO-7491 Trondheim, Norway}

\author{Oskar Kohout}
\affiliation{Institute for Quantum Optics and Quantum Information (IQOQI), Austrian Academy of Sciences, Boltzmanngasse 3, 1090 Vienna, Austria}
\affiliation{Fraunhofer Institute for Applied Optics and Precision Engineering IOF,Albert-Einstein-Strasse 7, 07745 Jena, Germany}
\affiliation{Friedrich-Schiller-Universität Jena FSU, Fürstengraben 1, 07743 Jena, Germany}

\author{Jan Lang}
\affiliation{Institute for Quantum Optics and Quantum Information (IQOQI), Austrian Academy of Sciences, Boltzmanngasse 3, 1090 Vienna, Austria}
%\affiliation{Vienna Center for Quantum Science and Technology (VCQ), Faculty of Physics, University of Vienna, Boltzmanngasse 5, 1090 Vienna, Austria}

\author{Sebastian Ecker}
\affiliation{Institute for Quantum Optics and Quantum Information (IQOQI), Austrian Academy of Sciences, Boltzmanngasse 3, 1090 Vienna, Austria}
\affiliation{Vienna Center for Quantum Science and Technology (VCQ), Faculty of Physics, University of Vienna, Boltzmanngasse 5, 1090 Vienna, Austria}

\author{Sebastian P. Neumann}
\affiliation{Institute for Quantum Optics and Quantum Information (IQOQI), Austrian Academy of Sciences, Boltzmanngasse 3, 1090 Vienna, Austria}
\affiliation{Vienna Center for Quantum Science and Technology (VCQ), Faculty of Physics, University of Vienna, Boltzmanngasse 5, 1090 Vienna, Austria}

\author{Julius Bittermann}
\affiliation{Institute for Quantum Optics and Quantum Information (IQOQI), Austrian Academy of Sciences, Boltzmanngasse 3, 1090 Vienna, Austria}
\affiliation{Vienna Center for Quantum Science and Technology (VCQ), Faculty of Physics, University of Vienna, Boltzmanngasse 5, 1090 Vienna, Austria}

\author{Robert Kindler}
\affiliation{Institute for Quantum Optics and Quantum Information (IQOQI), Austrian Academy of Sciences, Boltzmanngasse 3, 1090 Vienna, Austria}
\affiliation{Vienna Center for Quantum Science and Technology (VCQ), Faculty of Physics, University of Vienna, Boltzmanngasse 5, 1090 Vienna, Austria}

\author{Marcus Huber}
\email{marcus.huber@tuwien.ac.at}
\affiliation{Vienna Center for Quantum Science and Technology (VCQ), Atominstitut, Technische  Universit{\"a}t  Wien,  Stadionallee 2, 1020  Vienna,  Austria}
\affiliation{Institute for Quantum Optics and Quantum Information (IQOQI), Austrian Academy of Sciences, Boltzmanngasse 3, 1090 Vienna, Austria}

\author{Martin Bohmann}
\email{martin.bohmann@qtlabs.at}
\affiliation{Institute for Quantum Optics and Quantum Information (IQOQI), Austrian Academy of Sciences, Boltzmanngasse 3, 1090 Vienna, Austria}
\affiliation{Vienna Center for Quantum Science and Technology (VCQ), Faculty of Physics, University of Vienna, Boltzmanngasse 5, 1090 Vienna, Austria}
\affiliation{Quantum Technology Laboratories GmbH -- qtlabs, Clemens-Holzmeister-Straße 6/6, 1100 Vienna, Austria}

\author{Rupert Ursin}
\email{rupert.ursin@oeaw.ac.at}
\affiliation{Institute for Quantum Optics and Quantum Information (IQOQI), Austrian Academy of Sciences, Boltzmanngasse 3, 1090 Vienna, Austria}
\affiliation{Vienna Center for Quantum Science and Technology (VCQ), Faculty of Physics, University of Vienna, Boltzmanngasse 5, 1090 Vienna, Austria}

\author{Matej Pivoluska}
\email{mpivoluska@mail.muni.cz}
\affiliation{Vienna Center for Quantum Science and Technology (VCQ), Atominstitut, Technische  Universit{\"a}t  Wien,  Stadionallee 2, 1020  Vienna,  Austria}
\affiliation{Institute of Computer Science, Masaryk University, 602 00 Brno, Czech Republic}
\affiliation{Institute of Physics, Slovak Academy of Sciences, 845 11 Bratislava, Slovakia}
\affiliation{Institute for Quantum Optics and Quantum Information (IQOQI), Austrian Academy of Sciences, Boltzmanngasse 3, 1090 Vienna, Austria}

\date{\today}

\begin{abstract}

In a recent quantum key distribution experiment, high-dimensional protocols were used to show an improved noise resistance over a 10.2~km free-space channel. One of the unresolved questions in this context is whether the communicating parties actually shared genuine high-dimensional entanglement. In this letter we introduce an improved discretisation and entanglement certification scheme for high-dimensional time-bin setups and apply it to the data obtained during the experiment. Our analysis answers the aforementioned question affirmatively and thus the experiment constitutes the first transmission of genuine high-dimensional entanglement in a single degree of freedom over a long-range free-space channel.
\end{abstract}

\maketitle

\textit{Introduction.}---Quantum entanglement \cite{horodecki2009,Friis2019} is arguably {one of} the most important phenomena in quantum physics. 

{While in the early days of quantum mechanics, entanglement raised the controversial question of the completeness of the theory \cite{EPR1935},
 today entanglement is revealed as an invaluable resource, enabling many important quantum communication protocols, such as quantum teleportation \cite{PhysRevLett.70.1895}, super-dense coding \cite{PhysRevLett.69.2881}, and quantum key distribution \cite{PhysRevLett.67.661}.}{}

{In many applied implementations of quantum communication protocols, photons entangled in the  polarization degree of freedom are favored due to their simple and well established way of manipulation and measurement.}
However, the process of spontaneous parametric downconversion (SPDC), which is used to generate polarization entanglement, natively produces photons also {entangled} in other photonic degrees of freedom (DOFs){; see, e.g., \cite{10.5555/1817101,Erhard2020,Anwar2021}}{}.
Recently, there has been an increased interest in building setups to control these DOFs \cite{kwiat1997,barreiro2005,PhysRevA.96.040303,martin2017quantifying,Bavaresco2018,HerreraValencia2020highdimensional,Ortega2021,Achatz2022}  with the goal to  access  higher-dimensional Hilbert spaces.
The possible advantages of higher-dimensional (qudit) over two-dimensional (qubit) photonic entanglement in quantum communication were already observed at the beginning of this century ---
{qudit entanglement was shown to enable}
higher key rates and better noise resistance in QKD \cite{cozzolino2019}, higher communication rates in super-dense coding \cite{PhysRevLett.83.648} and novel protocols like the quantum secret sharing \cite{PhysRevA.65.022304}.
While qudit entangled states admit a higher amount of noise before losing their entanglement completely, it is a challenge to {efficiently} use this effect in {experimental} setups \cite{ecker2019overcoming,Hu2020} and even more so to harness this noisy entanglement in communication protocols \cite{doda2021quantum, Hu2021}. 
The reason for this is that in experiments over very noisy channels, such as free-space communication \cite{ursin2007}, satellite to ground links \cite{liao2017satellite} or underwater communication \cite{Feng:21,https://doi.org/10.48550/arxiv.2004.04821}, the surviving entanglement between two qudits is restricted to qubit subspaces and thus not genuinely high-dimensionally entangled \cite{ecker2019overcoming}.
Indeed, genuine high-dimensional entanglement has been certified before only for moderate distances and additionally, the experiment required the use of two different DOFs simultaneously \cite{Steinlechner2017}.

Recently, a high-dimensional QKD protocol was shown to yield a noise resilience advantage over a $10.2$~km metropolitan {horizontal} free-space channel \cite{HiDimQKD}.
Curiously, increased noise resistance is not a proof of genuine high-dimensional entanglement by itself. 
Indeed, to experience increased noise resistance, one can imagine a situation, where high-dimensional entanglement is created at the source, but the communicating parties at the end points can only certify entanglement in two dimensional subspaces --- this would be fully sufficient for QKD and the high-dimensional entanglement would therefore only serve as a resource to provide a noise-resistant distribution of qubit entanglement.
This would, however, always limit the rate to be below one secret bit per photon pair. 
The recent experiment \cite{HiDimQKD}, indeed showed strictly less key rate per coincidence, yet exhibited significant noise advantages. 
A natural follow-up question for understanding the role of high-dimensional entanglement in this context is whether both sides actually share a genuinely high-dimensional entangled state after the noisy free-space channel is applied to the high-dimensional states. 
In this letter we reanalyze the data from a new perspective and develop novel methods for certification of genuine high-dimensional time-bin entanglement. 
As a result,we report the distribution of genuine $3$-dimensional entanglement in a single DOF over a $10.2$~km metropolitan horizontal free-space channel.

\begin{figure*}[ht]
        \centering
        \includegraphics[width=2\columnwidth]{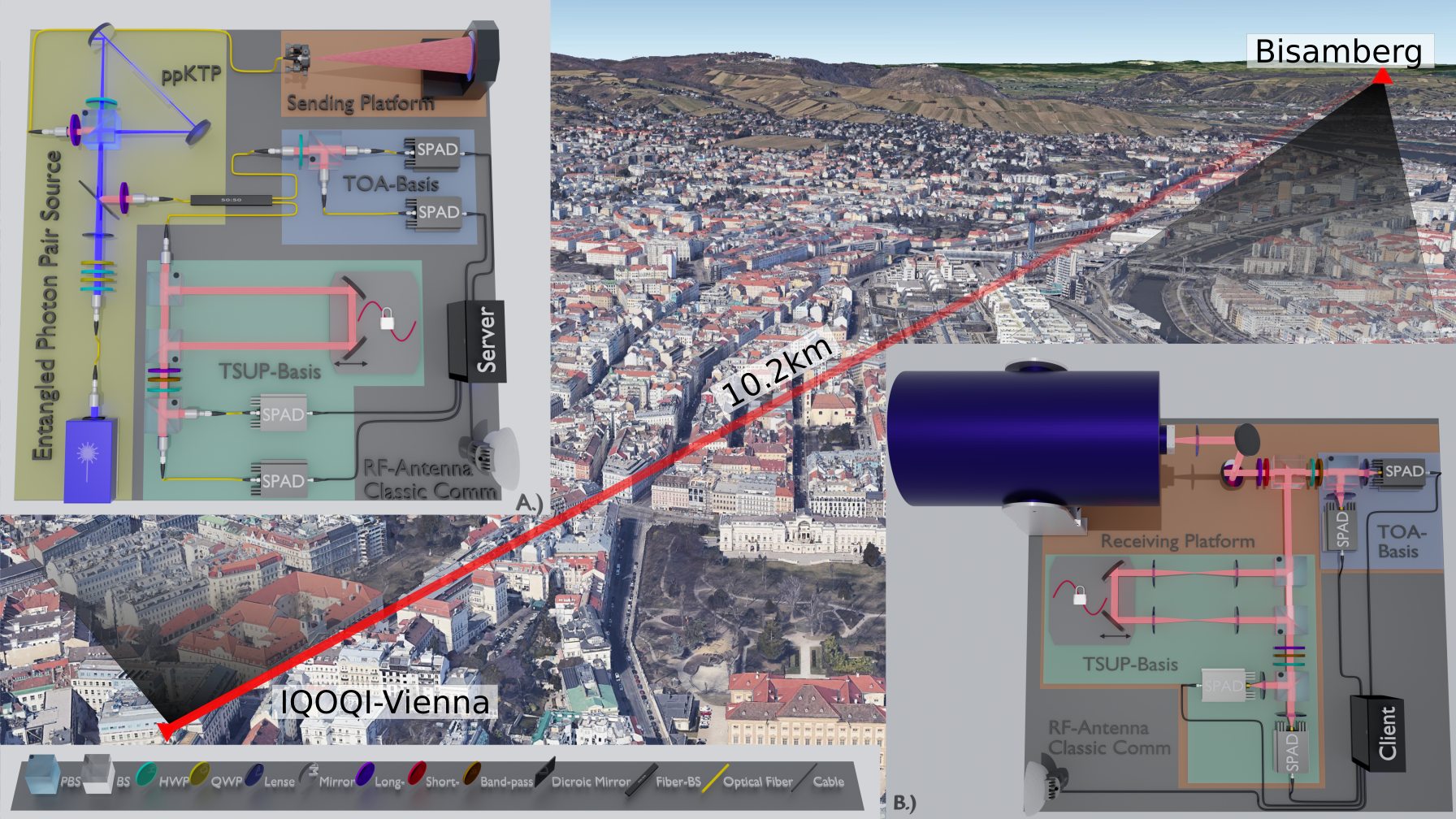}
        \caption{
        \textbf{{A.)}} The setup at the IQOQI laboratory (Alice) with a hyperentangled photon pair source pumped by a $^{39}K$ stabilized Laser at $404.532$ nm; 
        The hyperentangled state was set by adjusting a combination of half-wave plate (HWP) and quater-wave plates (QWP); The hyperentangled photons produced at $808.9$ nm in a ppKTP crystal are separated and guided to Alice and Bob. A 50:50 beam splitter (BS), at Alice's receiver, randomly sends the incoming photons to either to the basis measurement or the temporal superposition (TSUP) basis measurement; The Mach-Zehnder-Interferometer (MZI) in the TSUP basis is using Polarizing beam splitters followed by a measurement in D/A-basis.
        The photon detection events are recorded by single-photon avalanche diodes (SPAD), time stamped by the time-tagging module and streamed to the server.
        \textbf{B.)} Receiver at the Bisamberg laboratory (Bob). The received photons are guided to the detection modules after filtering the background noise with filter combinations.
        Bob's setup is identical with that of Alice, with the difference of having two 4f-systems in the MZI to compensate for the atmospheric turbulence and Bobs MZI is phase locked to a $^{87}Rb$ stabilized laser at $780.23$ nm; Both MZI's together form a post selection free Franson interferometer where each side is separated by 10.2~km.
        }
        \label{fig:linkv1}
    \end{figure*}

\textit{Experimental setup.}---
The experimental {photon-pair}{} state is created via {SPDC} by superposing wavelength-degenerate $\lambda_{\mathrm{SPDC}}=808.9\,$nm SPDC photons in a Sangnac interferometer. 
The SPDC photons are emitted by a type II ppKTP crystal, which is bidirectionally pumped with a $\lambda_{p}=404.453$~nm continuous wave laser.
The produced state is entangled simultaneously in the time-energy and the polarization DOFs and can be described as: 
 \begin{align}\label{eq:state}
          \ket{\Psi}_{\mathrm{AB}} =\int dt\, f(t) \ket{t,t}
          \otimes
          \left( \ket{H,H} +e^{-i\phi} \ket{V,V}\right),
    \end{align}
where $f(t)$ is a continuous function of time $t$, determined by the coherence time of the laser and $H(V)$ indicates the horizontal (vertical) polarization state. 
 The {generated} entangled photon pairs exit either side of the polarizing beam splitter of the Sagnac source; see Fig. \ref{fig:linkv1}.
 Then one photon is detected locally at Alice's lab where the source is located and the other one is guided to the sending platform and transmitted over a horizontal free-space channel to Bob's lab at $10.2$~km distance. 
 The sending platform consists of a Newtonian telescope with a mounted achromatic sending lens ($f=257\,$mm), where the former is used for tracking and the lens as a sending aperture.
 A Cassegrain telescope  ($f=2032\,$mm) receives the photons at Bob's lab and  guides them to the measurement setup.

Alice's and Bob's measurement setups consist of a time-of-arrival (TOA) and a time-superposition measurement (TSUP), each implementing measurement bases described in section \textit{Discretisation and certification}. 
The TOA-basis measurement consists of a projective measurement in the HV-basis and recording the photon arrival times in continuously streamed time-tags.
When processing these streams, one benefits from the hyperentangled state \cite{kwiat1997,barreiro2005} by taking polarization entanglement into account and passively sorting out photons not correlated in the polarization DOF, increasing the signal-to-noise ratio (SNR) in the TOA-basis.
The TSUP-basis uses an imbalanced Mach-Zehnder interferometer (MZI), with a path length difference of $\tau_{\mathrm{MZI}} = 2.7\,$ns, to superpose the photons' long and short path. 
The path difference $\tau_{\mathrm{MZI}}$ supersedes the coherence time of the SPDC photons ($\tau_{\mathrm{p}} \sim 3\,$ps), which is necessary to resolve the time of arrival of the detected photons. 
Alice's and Bob's local MZIs together form a Franson interferometer \cite{franson1989bell} to measure non-local two-photon interference.
The local MZIs consist of two polarizing beam splitters, which map the polarization horizontal/vertical (H/V)to the short/long path. 
Using a diagonal/anti-diagonal-polarization (D/A) measurement after the MZIs, one deletes the which-path information and completes the Franson interferometer for non-local temporal interferometry. 
By using the hyperentangled state, we increase the efficiency of the Franson interferometer by mapping polarization to time-energy entanglement (see Ref.\cite{Steinlechner2017} for details). 
Additionally, we implemented a $4f$-system {(see Ref.\cite{jin2018demonstration} for details)}{} in Bob's local MZI to compensate for the turbulent atmosphere.
Furthermore, both MZIs and the source have to be phase stable with respect to each other, which is achieved by phase-locking them onto an atomic hyper-fine transitions.
Further details on the experiment can be found in Ref. \cite{HiDimQKD},
where we have used the same free-space link to implement a noise-resistant high-dimensional QKD protocol. 

\textit{Results.}---In order to certify $3$-dimensional entanglement, we first
discretise the time-energy part of the produced continuous DOF state presented in Eq.\eqref{eq:state} into a $4\times 4$-dimensional state $\rho_{\Delta t}$.
This is achieved by coarse-graining the continuous time of arrival into time-bins of variable length $\Delta t$ (see \textit{Discretisation  and  certification} section for details and \cite{ecker2019overcoming,HiDimQKD}).
Then, for various choices $\Delta t$ we lower-bound the fidelity of $\rho_{\Delta t}$\ to the $4$-dimensional maximally entangled state $\ket{\phi_4^+} = \frac{1}{2} \sum_{i=0}^{3} \ket{ii}$. 
Fidelity values larger than $\frac{1}{2}$ certify the Schmidt number of $\rho_{\Delta t}$ (i.e. the smallest basis required to write down $\rho$) to be at least $3$ \cite{Bavaresco2018} and thus it certifies $\rho_{\Delta t}$ to be at least genuinely $3$-dimensionally entangled.

\begin{figure}[h!]
    \centering
    \includegraphics[width=\columnwidth]{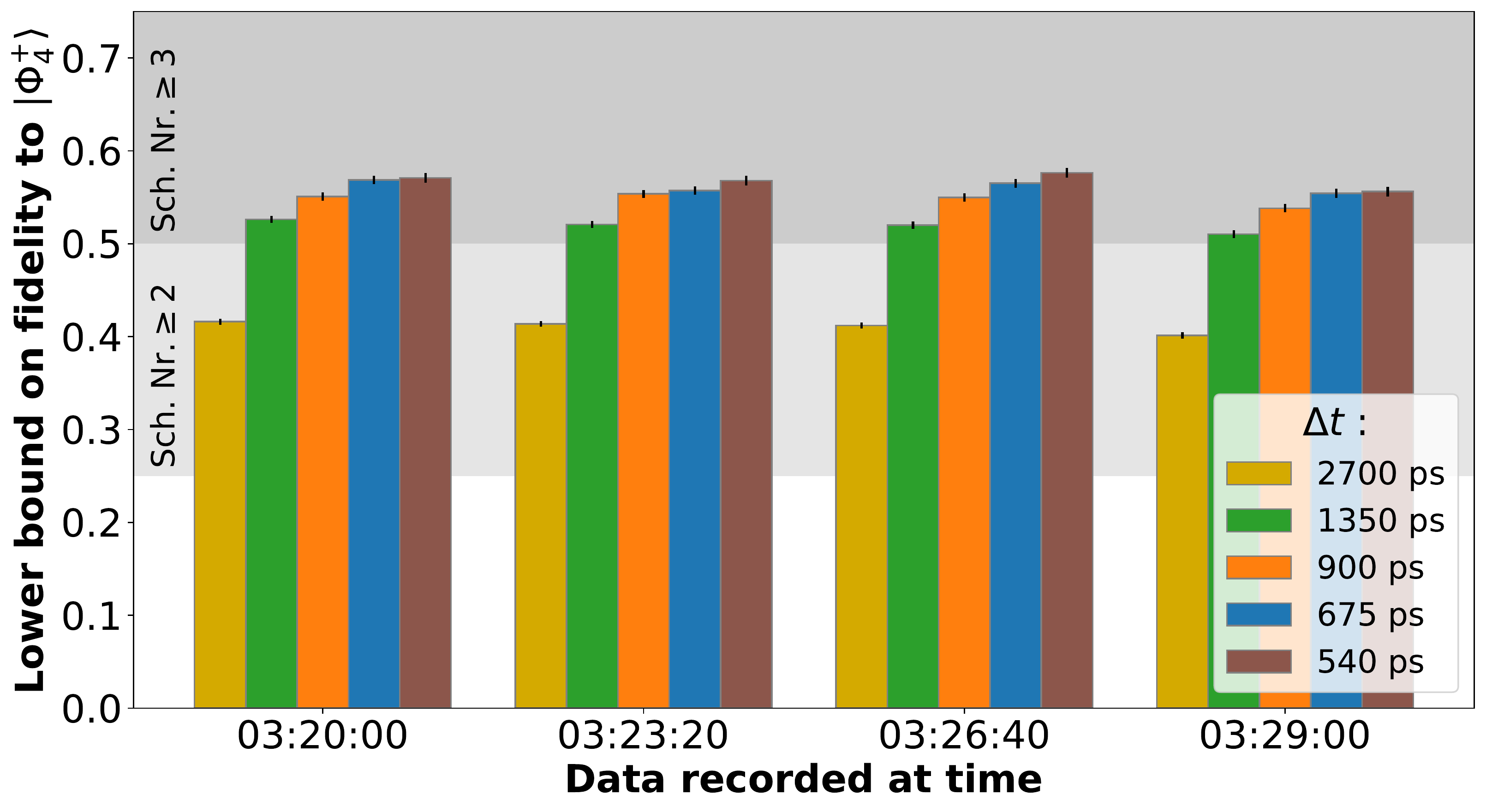}
    \caption{ 
    \textbf{Entanglement dimensionality witness.} In order to certify entanglement dimensionality we use different discretisations with $d=4$. 
    For different time-bin length $\Delta t$, we evaluate the fidelity of the time-energy part of the discretised distributed state to maximally entangled state
    $|\phi^{+}_4\rangle=1/\sqrt{4}\sum^{3}_{i=0}|ii\rangle$. 
    The integration time of all data points is $200$~s and the fidelity is evaluated throughout the duration of the experiment.
    Fidelity strictly above $0.25$ certifies that the underlying state has Schmidt number of at least $2$ and thus is entangled, while fidelity strictly above $0.5$ certifies that the underlying state has Schmidt number value of at least $3$ and thus the correlations present in the state cannot be encoded using any qubit entangled state.
    }
    \label{fig:qutrit_entanglement}
\end{figure}

Lower bounds on fidelity graphically presented in FIG. \ref{fig:qutrit_entanglement} were certified for $200$s blocks of discretised measurement data obtained at different points in time during the experiment. 
The results clearly show that the certified fidelity is higher for shorter time-bins and  the highest fidelity is observed at $\Delta t = 540$~ps.  
{This observation can be explained by the random distribution of accidental coincidence counts caused by external noise. In contrast, the signal coincidences caused by the entangled photons are always time-correlated in the time of arrival. Therefore, choosing shorter time-bins in effect filters out more accidental counts. 
The minimal size of $\Delta t$ is fundamentally limited by the coherence time of the SPDC photons $\tau_{\mathrm{p}}$. Further, feasible values of $\Delta t$ are severely restricted by the electronic jitter of the detectors and the time-tagging modules. 
All of these effects decrease the precision of time-tagging. 
This kind of noise affects the certified fidelity stronger for lower values of $\Delta t$. 
We have indeed observed that for $\Delta t$ below $540\,$ps the certified fidelity starts to rapidly decrease again.
}

\textit{Discretisation and certification.}---
The crucial component of TOA discretisation is the coarse-graining of time-tagged detection event streams into time-bins of length $\Delta t$.
Subsequently, $d$ time-bins are collected into a time-frame, which determines the dimensionality $d\times d$ of the discretised Hilbert space.
Conventionally, time-frames are composed of $d$ subsequent time-bins (see e.g. \cite{ecker2019overcoming,HiDimQKD}). 
Applying this approach to our data does not lead to a certification of high-dimensional entanglement (see FIG. \ref{fig:qutrit_entanglement} with $\Delta t = 2700$~ps).
Intuitively, with $\Delta t = 2700$~ps one observes a large proportion of time-frames with more than a single detection event on either Alice's or Bob’s side. To not incur additional assumptions, these are assigned random outcomes, which degrade the data quality significantly. 
In order to have shorter time-frames it is instrumental to be able to decrease the time-bin length $\Delta t$.
For the conventional method this requirement translates to shortening the MZI delay $\tau_{\mathrm{MZI}}$ on demand (see Ref~\cite{martin2017quantifying} for an experiment with variable $\tau_{\mathrm{MZI}}$), or use multiple MZI interferometers. However, both methods are hard to implement for spatially separated interferometers.

\begin{figure}
    \centering
    \includegraphics[width = \columnwidth]{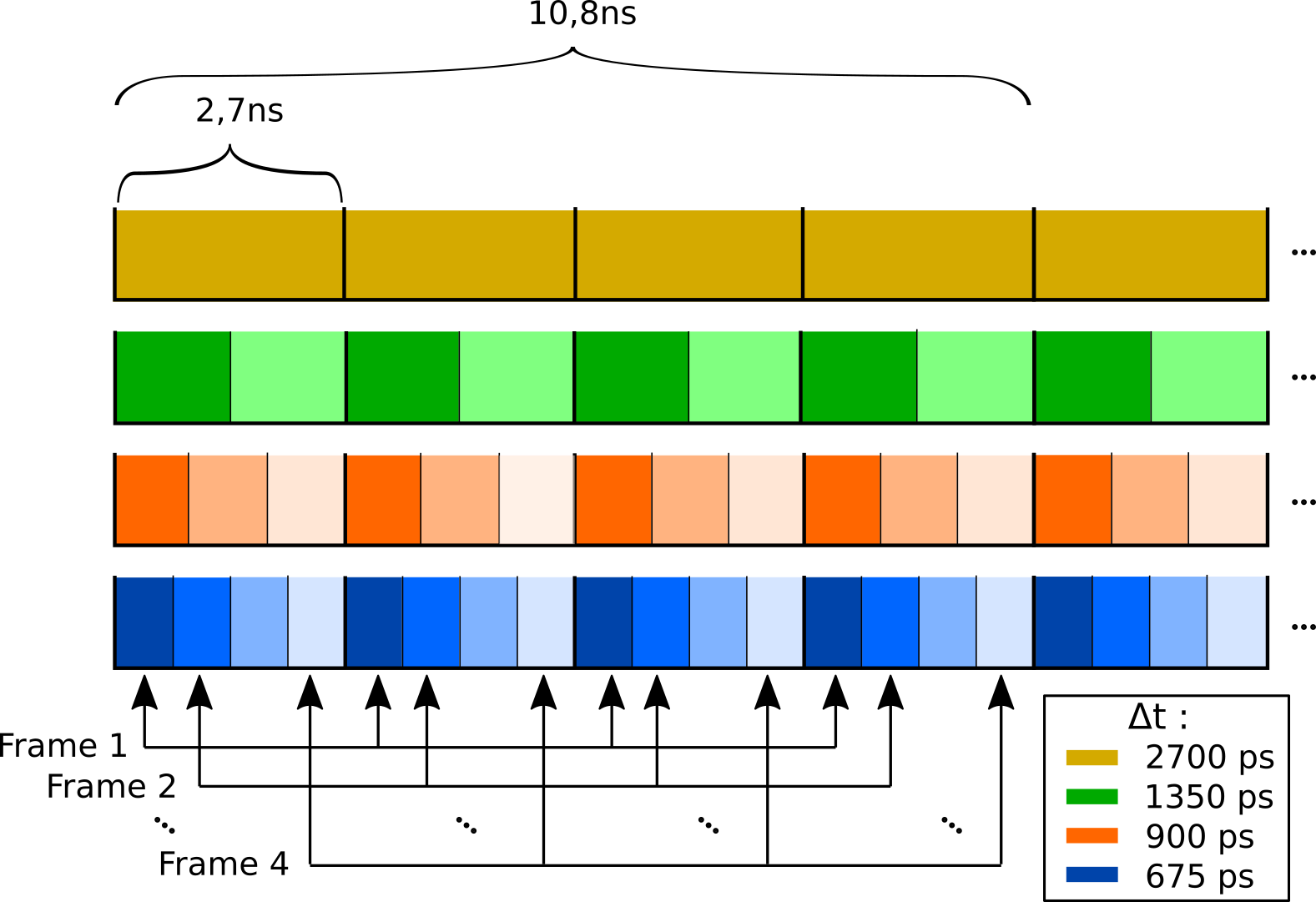}
    \caption{Discretisation we used to calculate the Schmidt number certificates. The time delay between interferometer arms is $\tau_{\mathrm{MZI}}=2.7$~ns.Each measured time interval of length $10.8$~ns is divided into time-bins of size $\Delta t$. 
    Subsequently, time-bins in the analyzed time-interval with starting times exactly $2.7$~ns apart are collected into a time frame. These were chosen specifically, because we can use our measurement setup to measure projections onto some superpositions of bins with the same color.
    }
    \label{fig:discretisation}
\end{figure}

In order to overcome this complication, we utilize a novel coarse-graining scheme, which allows us to probe $4$-dimensional subspaces spanned by non-neighboring time-bins of variable length $\Delta t$.
Our method assigns $4$ time-bins with starting times exactly $\tau_\mathrm{MZI}$ apart to a single time-frame. 
If the time-bin length $\Delta t$ divides $\tau_\mathrm{MZI}$, such time-frames can be interleaved so that a whole time-tagged measurement stream is covered (see FIG. \ref{fig:discretisation}). 
The main advantage of this method is that time-bins probing $4$ dimensional subspaces can be much shorter than required by the conventional method. 
Such short time-frames allow for a larger fraction of accidental clicks to be post-selected away as single clicks.

Subsequently, we associate basis vectors $\ket{0},\ket{1},\ket{2},\ket{3}$ to time-bins in each time-frame.
The TOA measurement therefore corresponds to a projection onto the computational basis of the $4\times 4$ Hilbert space defined by the time-bin length $\Delta t$.
Definition of the discretised TSUP measurement is a little more involved. 
The Franson interferometer  implemented in the TSUP measurement modules can essentially be seen as implementing projections onto superpositions of two time-bins, the first starting at time $t$ and the second one starting at time $t+\tau_\mathrm{MZI}$.
For time-frames we defined above this corresponds to projections onto six vectors: $\left\{\frac{1}{\sqrt{2}}(\ket{0}\pm\ket{1}),\frac{1}{\sqrt{2}}(\ket{1}\pm\ket{2}),\frac{1}{\sqrt{2}}(\ket{2}\pm\ket{3})\right\}$.
Clearly $\left\{\frac{1}{\sqrt{2}}(\ket{0}\pm\ket{1}),\frac{1}{\sqrt{2}}(\ket{2}\pm\ket{3})\right\}$ form a complete $4$-dimensional Hilbert space basis and we will denote the projective measurement associated to this basis $\mathrm{TSUP}_1$.
On the other hand, vectors $\frac{1}{\sqrt{2}}(\ket{1}\pm\ket{2})$ need to be supplemented with another two orthogonal vectors from a subspace spanned by $\ket{0}$ and $\ket{3}$. 
We opted to approximate measurement counts for projection onto this subspace with counts for projections onto $\left\{\frac{1}{\sqrt{2}}(\ket{3}\pm\ket{0'})\right\}$, where $\ket{0'}$ is a time-bin starting exactly $\tau_\mathrm{MZI}$ after the start of time-bin associated with $\ket{3}$. We denote the projective measurement associated to this second basis as TSUP$_2$.
Coincidence matrices for simultaneous measurements by both parties in these three bases can be obtained from the measurement data and post-processed into outcome probabilities.
These are subsequently used to obtain lower bounds for density matrix elements $\bra{ii}\rho_{\Delta t}\ket{ii}$ for all $i\in\{0,1,2,3\}$ and for $\Re(\bra{ii}\rho_{\Delta t}\ket{i+1,i+1}) = \Re(\bra{i+1,i+1}\rho_{\Delta t}\ket{i,i})$ for $i\in\{0,1,2\}$.
{While the former is just a probability to obtain the $i^\mathrm{th}$ outcome in Alice's and Bob's lab simulatneously, the latter can be obtained using the following calculation:
\begin{align}\nonumber
    &\tfrac{(\bra{i}+\bra{i+1})_A\otimes (\bra{i}+\bra{i+1})_B\,\rho_{\Delta t}\, (\ket{i}+\ket{i+1})_A\otimes (\ket{i}+\ket{i+1})_B}{4}\\\nonumber
    +&\,\tfrac{(\bra{i}-\bra{i+1})_A\otimes (\bra{i}-\bra{i+1})_B\,\rho_{\Delta t}\, (\ket{i}-\ket{i+1})_A\otimes (\ket{i}-\ket{i+1})_B}{4}\\\nonumber
    -&\,\tfrac{(\bra{i}-\bra{i+1})_A\otimes (\bra{i}+\bra{i+1})_B\,\rho_{\Delta t}\, (\ket{i}-\ket{i+1})_A\otimes (\ket{i}+\ket{i+1})_B}{4}\\\nonumber
    -&\,\tfrac{(\bra{i}+\bra{i+1})_A\otimes (\bra{i}-\bra{i+1})_B\,\rho_{\Delta t}\, (\ket{i}+\ket{i+1})_A\otimes (\ket{i}-\ket{i+1})_B}{4} \\
    =&\,2\Re[\bra{i,i}\rho_{\Delta t}\ket{i\!+\!1,i\!+\!1}]\!+\! 2\Re[\bra{i,i\!+\!1}\rho_{\Delta t}\ket{i\!+\!1,i}]. \label{eq:estimation1}
\end{align}
In order to obtain a lower bound on $2\Re[\bra{i,i}\rho_{\Delta t}\ket{i+1,i+1}]$ it remains to derive an upper bound on the term $2\Re[\bra{i,i+1}\rho_{\Delta t}\ket{i+1,i}]$.
This can be done as follows.
First, note that the real part of a complex number is always smaller than its absolute value, therefore 
\begin{align}\nonumber
    2&\Re[\bra{i,i+1}\rho_{\Delta t}\ket{i+1,i}] \\\nonumber
    &\leq 2|\bra{i,i+1}\rho_{\Delta t}\ket{i+1,i}| \\
    & \leq 2\sqrt{\bra{i,i+1}\rho_{\Delta t}\ket{i,i+1}\bra{i+1,i}\rho_{\Delta t}\ket{i+1,i}}, \label{eq:estimation2}
\end{align}
where the last inequality is an application of the Cauchy-Schwartz inequality: $$\vert\bra{ij}\rho\ket{kl}\vert \leq \sqrt{\bra{ij}\rho\ket{ij}\bra{kl}\rho\ket{kl}}.$$}{}
Note that the fidelity of $\rho_{\Delta t}$ to $\ket{\phi^+_4}$ is defined as
\begin{equation}
  F(\rho_{\Delta t},\ket{\phi^{+}_4}\bra{\phi^{+}_4})=   \frac{1}{4} \sum_{i,j=0}^3 \Re(\bra{ii}\rho_{\Delta t}\ket{jj}),
\end{equation} 
and thus it remains to lower bound the six remaining real parts of density matrix elements $\Re(\bra{ii}\rho_{\Delta t}\ket{jj})$, with $|i-j| > 1$.
This can be done using density matrix completion techniques based on the Sylvester’s criterion \cite{PhysRevA.96.040303,martin2017quantifying}, from which it follows that every subdeterminant of a positive
semidefinite matrix should be non-negative.
Performing these calculations on our experimental data leads to fidelity lower bounds which are presented in FIG.\ref{fig:qutrit_entanglement}.

\textit{Discussion.}--- {In this letter we revisit a recent experiment, where high-dimensional QKD experiment was performed over a long-distance free-space channel to successfully show an advantage in noise-resistance.
We shine more light on the role high-dimensional entanglement has played in this experiment. 
In particular, it was not clear whether the genuine high-dimensional entanglement created at the source actually survives the transmission over a very noisy atmospheric channel.
Here we answer this question affirmatively. 
This was possible only thanks to developing a novel way of looking at time-bin entanglement, by relaxing the requirement that a time-frame is composed of chronologically subsequent time-bins. 
Indeed with the new technique we were able to use much shorter time-bins for supreme noise filtering and thus
report an experimental realization of high-dimensional entanglement distribution over a long-distance horizontal free-space channel. 
Such a feat, to the best of our knowledge, has never been achieved before in a single DOF.
We also believe that the techniques developed within our research can be further improved on, which would allow us to certify even higher-dimensional entanglement in the time-energy DOF. 
The simplest and most immediate upgrade to the setup would be adding another measurement setting to the Franson interferometer setup, which should allow us to add phase to the photon passing through the long arm and thus be able to measure projections on $\tfrac{\ket{k}\pm i\ket{k+1}}{\sqrt{2}}$. 
This would in effect improve our estimation of the first off-diagonal density matrix element (Eq.\eqref{eq:estimation1}) so that the estimation performed in Eq.\eqref{eq:estimation2} would not be needed. 
The second experimental improvement is more involved and consists of adding a capability to modify the MZI delay $\tau_{MZI}$. 
Already measurements with two different values of $\tau_{MZI}$ as in \cite{martin2017quantifying} would allow efficient estimation of a larger number of density matrix elements of the experimental state. 
This would allow more stringent constrains for the  semi-definite programs used to complete the density matrix in the last step of the analysis, thus increasing the tightness of our bound.
}{}

\textit{Acknowledgements.}---
M.~P. acknowledges funding  from GAMU project MUNI/G/1596/2019 and VEGA project No. 2/0136/19.
M.~H. would like to
acknowledge funding from the European Research Council
(Consolidator grant ’Cocoquest’ 101043705).
M.~P. and M.~H. acknowledge funding from European Commission (grant 'Hyperspace' 101070168).
L.~B. acknowledges European Union’s Horizon 2020 programme grant  agreement  No.857156  (OpenQKD)  and  the Austrian  Academy of  Sciences  in  cooperation  with  the FhG  ICON-Programm  “Integrated  Photonic  Solutionsfor Quantum Technologies (InteQuant)”. 
L.~B. also gratefully  acknowledges  financial  support  from  the  Austrian Research  Promotion  Agency  (FFG)  Agentur  f\"ur  Luft-und Raumfahrt (FFG-ALR contract 844360 and 854022).

\end{document}